\RequirePackage[2020-02-02]{latexrelease}
\documentclass[aps,prb,twocolumn,showpacs,superscriptaddress,floatfix,longbibliography]{revtex4-2}
\pdfoutput=1
\usepackage[titletoc,toc,title]{appendix}
\usepackage{amsmath,amssymb}
\usepackage{graphicx}
\usepackage{color}
\usepackage{rotating}
\usepackage{bm}
\usepackage{setspace}
\usepackage{hyperref}
\usepackage{float}
\usepackage{soul}
\usepackage[T1]{fontenc}
\usepackage{bbold}
\usepackage{color}
\usepackage{braket}
\usepackage{physics}
\usepackage{subcaption}

\graphicspath{{figures/}}

\hypersetup{
colorlinks=true,
urlcolor= blue,
citecolor=blue,
linkcolor= blue,
bookmarks=true,
bookmarksopen=false,
}

\renewcommand{\vec}[1]{\bm{#1}}

\begin{document}

\preprint{APS/123-QED}

\title{Electric field-tunable layer polarization in  graphene/boron nitride twisted quadrilayer superlattices}

\author{Ziyan Zhu}
\altaffiliation{Present address: Stanford Institute for Materials and Energy Sciences,
SLAC National Accelerator Laboratory, Menlo Park, CA 94025, USA}
\email{ziyanzhu@stanford.edu}
\affiliation{Department of Physics, Harvard University, Cambridge, Massachusetts 02138, USA}

\author{Stephen Carr}
\affiliation{Brown Theoretical Physics Center and Department of Physics, Brown University, Providence, RI 02912, USA}
\author{Qiong Ma}
\affiliation{Department of Physics, Boston College, Chestnut Hill, MA 02467, USA}
\affiliation{CIFAR Azrieli Global Scholars program, CIFAR, Toronto, Canada}
\author{Efthimios Kaxiras}
\affiliation{Department of Physics, Harvard University, Cambridge, Massachusetts 02138, USA}
\affiliation{John A. Paulson School of Engineering and Applied Sciences, Harvard University, Cambridge, Massachusetts 02138, USA}

\begin{abstract}
      The recently observed unconventional ferroelectricity in AB bilayer graphene sandwiched by hexagonal Boron Nitride (hBN) presents a new platform to manipulate correlated phases in multilayered van der Waals heterostructures~\citep{zheng2020}. 
      We present a low-energy continuum model for AB bilayer graphene encapsulated by the top and bottom layers of either hBN or graphene, with two independent twist angles. For the graphene/hBN heterostructures, we show that twist angle asymmetry leads to a layer polarization of the valence and conduction bands. We also show that an out-of-plane displacement field not only tunes the layer polarization but also flattens the low-energy bands. We extend the model to show that the electronic structures of quadrilayer graphene heterostructure consisting of AB bilayer graphene encapsulated by the top and bottom graphene layers can similarly be tuned by an external electric field. 
\end{abstract}
\maketitle

\section{Introduction}

Manipulating the twist angle or the lattice mismatch in stacks of multilayered two-dimensional (2D) materials, referred to as van der Waals (vdW) heterostructures, introduces a long-wavelength moir\'e potential that fundamentally alters the electronic properties of the constituent materials~\cite{carr2017twistronics}. A plethora of unconventional states has been observed in single-twist moir\'e vdW heterostructures including twisted bilayer graphene~\cite{cao2018correlated,cao2018unconventional,lu2019superconductors,cao2021nematicity,cao2021pauli}, transition metal dichalcogenides~\cite{wang2019wse2,regan2020mott,tang2020hubbard,jin2021stripes,xu2021coexisting,xu2022hubbard}, twisted double bilayer graphene~\cite{liu2020dblg,burg2019dblg,shen2020correlated,cao2020dblg}, alternatively twisted trilayer and multi-layered graphene~\cite{hao2021electric,park2021tunable,park2021magic}, and monolayer-bilayer graphene~\cite{he2021competing}. Following these observations, forays into multilayered vdW heterostructures with a higher order ``moir\'e of moir\'e'' pattern (also known as ``double moir\'e'' / ``super moir\'e''), such as twisted trilayer graphene with two independent twist angles, have led to discoveries of novel correlated states, among other mechanical and electronic properties~\cite{amorim2018electronic,mora2019flat,zhu2020relaxation,zhu2020ttlg,tsai2019correlated,turkel2021twistons}. In these systems, the second twist angle and/or lattice mismatch provides additional flexibility to tune the properties of 2D layered materials.

Moir\'e of moir\'e patterns are a consequence of the interference between two different bilayer moir\'e patterns, and they have length scales that are generally orders of magnitude longer than the first-order moir\'e length. 
As a result of the higher-order interference pattern, such a system lacks a periodic approximation even in the continuum limit, and consequently has no approximate Brillouin zone in reciprocal space ~\cite{andelkovic2019double,zhu2020relaxation,zhu2020ttlg,oka2021fractal}. The lack of periodicity and the large system size present significant challenges to the usual approaches of developing an accurate theoretical model. The electronic band structure of these moir\'e of moir\'e systems has been calculated using a momentum-space low-energy continuum model for twisted trilayer graphene~\cite{amorim2018electronic,zhu2020ttlg} and twisted hexagonal Boron Nitride (hBN)/monolayer graphene/hBN heterostructures~\cite{oka2021fractal,andelkovic2019double}. This approach is both computationally efficient and accurate for the low-energy features of interest. 

\begin{figure*}[ht!]
    \centering
    \includegraphics[width=\linewidth]{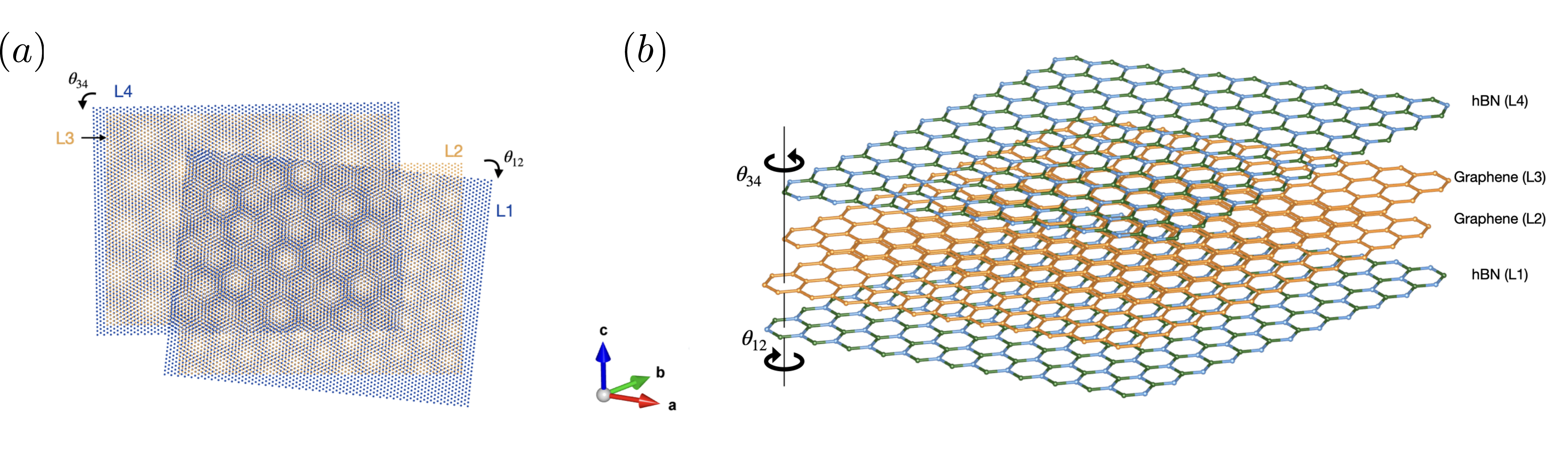}
    \caption{(a) Moir\'e patterns in hBN-encapsulated AB bilayer graphene. Blue and yellow corresponds to hBN and graphene layers respectively. L1 is twisted by $7^\circ$ clockwise ($\theta_{12} = 7^\circ$), and L4 is twisted $1^\circ$ counter-clockwise ($\theta_{34} = 1^\circ$). (b) Perspective view of hBN encapsulated AB bilayer graphene. }
    \label{fig:geom}
\end{figure*}

By encapsulating monolayer or Bernal (AB) bilayer graphene in hBN, a moir\'e of moir\'e pattern can be created from the interference between two bilayer moir\'e superlattices: one from the twist angle, and the other from the $\sim$1.7\% graphene-hBN lattice constant mismatch~\cite{andelkovic2019double,wang2019new,moriyama2019observation,finny2019tunable,wang2019composite,yang2020insitu,loconte2020commensurate,onodera2020cyclotron,chen2022first,oka2021fractal,zheng2020}.
In particular, unconventional ferroelectricity has been experimentally observed in heterostructures consisting of an AB bilayer graphene sandwiched between top and bottom hBN layers~\cite{zheng2020}, in contrast to the conventional ferroelectricity in other 2D vdW heterostructures due to a net structural polarization ~\cite{bune1998,chang2016SnTe,liu2016CuInP2S6,zhou2017in2se3,cui2018intercorrelated,yuan2019wete2,fei2018wte2,yasuda2021stacking,vizner2021interfacial}. In the graphene/hBN heterostructures, hysteretic loops in the top-gate/back-gate phase-space have been observed. The top-gate voltages corresponding to resistance peaks stay constant for a wide range of back-gate voltages, which is referred to as Layer Specific Anomalous Screening (LSAS). However, the ability to confirm the electronic origin of the observed ferroelectricity has been hindered by the lack of an accurate microscopic theoretical model. In this work we aim to provide a single-particle low-energy continuum model for this quadrilayer graphene/hBN heterostructures and provide a theoretical foundation to understand the observed ferroelectricity in this system. We find that the graphene/hBN heterostructure exhibits a moir\'e-induced layer polarization that sensitively depends on the twist angle. The layer polarization can be tuned by the application of an external displacement field. This moir\'e-induced layer polarization helps explain the observed unconventional ferroelectricity and LSAS. In addition, we extend the model to study a similar quadrilayer system, AB bilayer graphene encapsulated by top and bottom monolayer graphene, whose electronic structure is also electric field-tunable. 

The paper is organized as follows. In Section~\ref{sec:methods}, we discuss the system geometry and review the low-energy continuum model for moir\'e of moir\'e systems and generalize to twisted quadrilayer graphene/hBN heterostructures. In Section~\ref{sec:results}, we present the main results and show the electric field-tunable layer polarization of the valence and conduction bands. In Section~\ref{sec:graphene}, we extend our low-energy continuum to study the electronic structure of quadrilayer graphene heterostructures. We summarize our findings and make connections to experimental observations in Section~\ref{sec:conclusion}.

\section{Theoretical Model} \label{sec:methods}

\subsection{Geometry}
The real space geometry of the system is shown in Fig.~\ref{fig:geom}. The top layer, layer 1 (L1), and the bottom layer, layer 4 (L4), are twisted with respect to the sandwiched AB bilayer graphene by $\theta_{12}$ and $\theta_{34}$ respectively, with $\theta_{12}$ in the clockwise direction and $\theta_{34}$ in the counter-clockwise direction. The monolayer lattice constants are defined as the column vectors of the following matrix: 
\begin{equation}
    A_i = a_i \begin{pmatrix} 1  & 1/2 \\ 
    0 & \sqrt{3}/2
    \end{pmatrix} = \begin{pmatrix} \vec{a}_1 & \vec{a}_2
    \end{pmatrix},
\end{equation}
where $i$ = Gr/hBN, and $a_\mathrm{Gr} = 2.4768$ \AA \ and $a_\mathrm{hBN} = 2.5189$ \AA \  are the graphene and hBN lattice constants respectively.
Defining the counter-clockwise rotation matrix
\begin{equation}
\mathcal{R} (\theta) = 
    \begin{pmatrix} 
    \cos\theta & -\sin\theta \\ 
    \sin\theta & \cos\theta
    \end{pmatrix},\label{eqn:rotation}
\end{equation}
the lattice vectors of the 4 layers are given as $A_1 = \mathcal{R} (\theta_{12}) A_\mathrm{hBN}, A_2 = A_3 = A_\mathrm{Gr}, A_4 = \mathcal{R}(\theta_{34}) A_\mathrm{hBN}.$ The monolayer reciprocal lattice vectors are given by the column vectors of $G_\ell = 2\pi A_\ell^{-T}$ for $\ell = 1,2,3,4$. In terms of $G_\ell$, the bilayer moir\'e Brillouin zones are spanned by the column vectors of $G_{\ell, \ell+1} = G_{\ell+1}- G_\ell = 2\pi(A_{\ell+1}^{-T} - A_\ell^{-T}),$ where $\ell=1,3$ for the top or bottom Gr/hBN moir\'e interface respectively. The moir\'e supercell in real space between layers $l$ and $l+1$ is $A_{\ell,\ell+1} = 1/(2\pi) G_{\ell,\ell+1}^{-T}$. The norm of the column vectors are the moir\'e length due to the twist and lattice mismatch, which is given by
\begin{equation}
    \lambda_{\ell, \ell+1} = \frac{(1+\delta)a_\mathrm{Gr} }{\sqrt{2(1+\delta)(1-\cos\theta_{\ell,\ell+1}) + \delta^2 } },
\end{equation}
where $\delta = a_\mathrm{hBN} / a_\mathrm{Gr} = 1.017$ is the lattice constant mismatch ratio between graphene and hBN~\cite{yankowitz2012emergence}. Each moir\'e pattern of the two interfaces, L1/L2 and L3/L4, exhibits a single coherent moir\'e length (see the interference patterns between L1, L2 and L3, L4 in Fig.~\ref{fig:geom}(a)). These two moir\'e supercells have different length scales and are rotated relative to each other, and their interference pattern forms the more complex higher-order moir\'e of moir\'e patterns (Fig.~\ref{fig:geom}(a)). Figure~\ref{fig:geom}(b) shows a perspective view of the system. We assume that the top and bottom hBN layers are rotated by $180^\circ$ such that in the absence of the twist, the system has a mirror symmetry along the $z$-direction. 
With a twist angle, however, the stacking configuration varies in space and we expect all possible stackings to occur in the system, and as a result, the inversion symmetry is broken. 
We checked that when the top and bottom hBN layers are not rotated, the reported results have no qualitative difference, except for the slight asymmetry under the exchange of $\theta_{12}$ and $\theta_{34}$.

\begin{figure}[ht!]
    \centering
    \includegraphics[width=\linewidth]{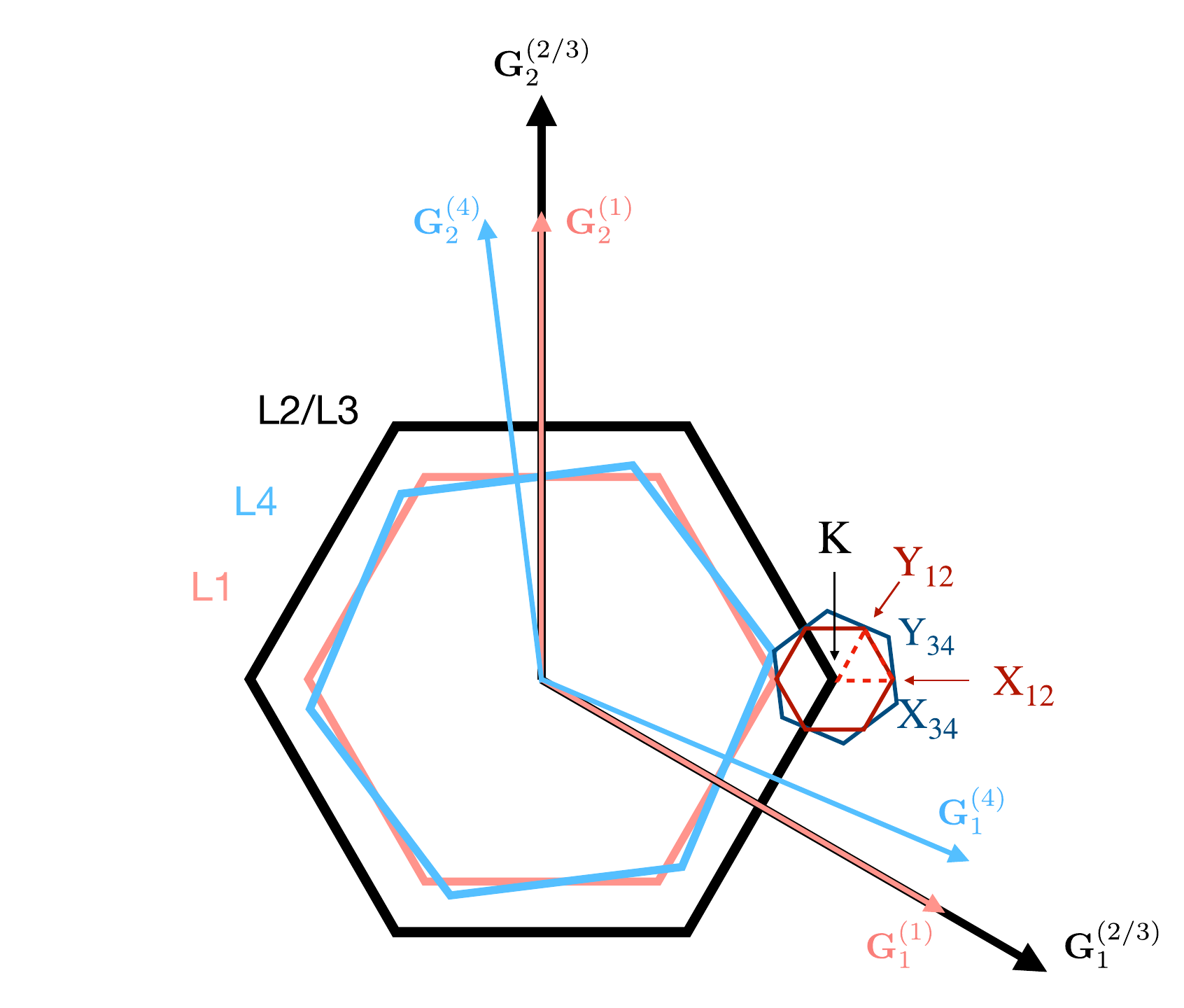}
    \caption{Illustration of Brillouin zone folding of twisted quadrilayer heterostructures with $\theta_{12}=0^\circ, \theta_{34}=7^\circ$. For visual clarity, the lattice constant ratio between L1 and L2 (also between L4 and L3) is taken to be a hypothetical value of 5/4. Light red, black, and light blue represent the monolayer Brillouin zones of L1, L2/L3, L4 respectively. The arrows are the monolayer primitive reciprocal lattice vectors. The small red and blue hexagons on the right are the moir\'e Brillouin zones that correspond to the L1/L2 and L3/L4 interfaces respectively. High symmetry points of the first ($X_{12}$, $Y_{12}$) and second interface ($X_{34}$, $Y_{34}$) moir\'e Brillouin zones, and the monolayer graphene Dirac point ($K$) are labeled.}
    \label{fig:reciprocal}
\end{figure}

Figure~\ref{fig:reciprocal} shows the reciprocal space of a hypothetical quadrilayer system, qualitatively similar to the graphene/hBN system. Note that due to broken sublattice symmetry, the Brillouin zone corners of the hBN monolayers (top and bottom layers) are $X$ and $Y$ and they are nonequivalent, unlike the $K$ and $K'$ points in graphene (middle layers). There are two moir\'e Brillouin zones that correspond to the top and bottom graphene-hBN interfaces (dark red and blue hexagons in Fig.~\ref{fig:reciprocal}). These two moir\'e Brillouin zones are incommensurate, and thus the system has no overall Brillouin zone. In the band structures of graphene/hBN heterostructures in the rest of the paper, we plot the energy eigenstates along the high-symmetry line in the moir\'e Brillouin zone that corresponds to the L1/L2 interface (red dashed lines in Fig.~\ref{fig:reciprocal}).

\begin{figure*}[ht!]
    \centering
    \includegraphics[width=0.8\linewidth]{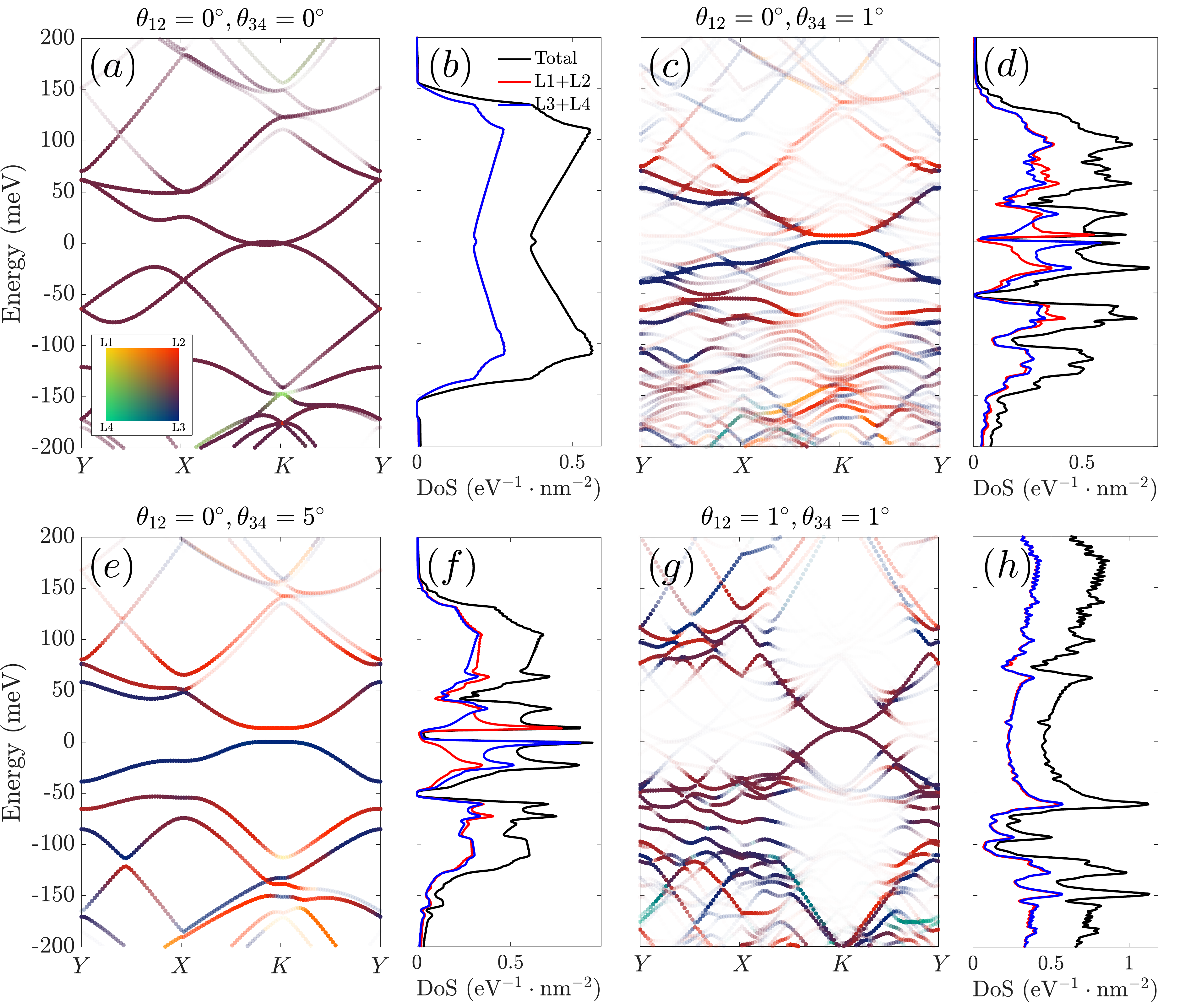}
    \caption{Band structure and corresponding density of states for the quadrilayer graphene-hBN heterostructure for (a)-(b) $\theta_{12} = 0^\circ, \theta_{34} = 0^\circ$ (c)-(d) $\theta_{12} = 0^\circ, \theta_{34} = 1^\circ$, (e)-(f) $\theta_{12} = 0^\circ, \theta_{34} = 5^\circ$, (g)-(h) $\theta_{12}= 1^\circ, \theta_{34}=1^\circ.$ Colors of the band structure represent the projection of the wavefunction weights onto the center site along the high symmetry line for each layer. The color scale is indicated in the panel on the bottom left side of (a), with yellow, red, blue, green corresponding to layers 1-4 respectively. Transparency is inversely proportional to the magnitude of the projected wavefunction weights. The high symmetry line is indicated by the red dashed lines in Fig.~\ref{fig:reciprocal}.} 
    \label{fig:bands}
\end{figure*}

\begin{figure}[ht!]
    \centering
    \includegraphics[width=\linewidth]{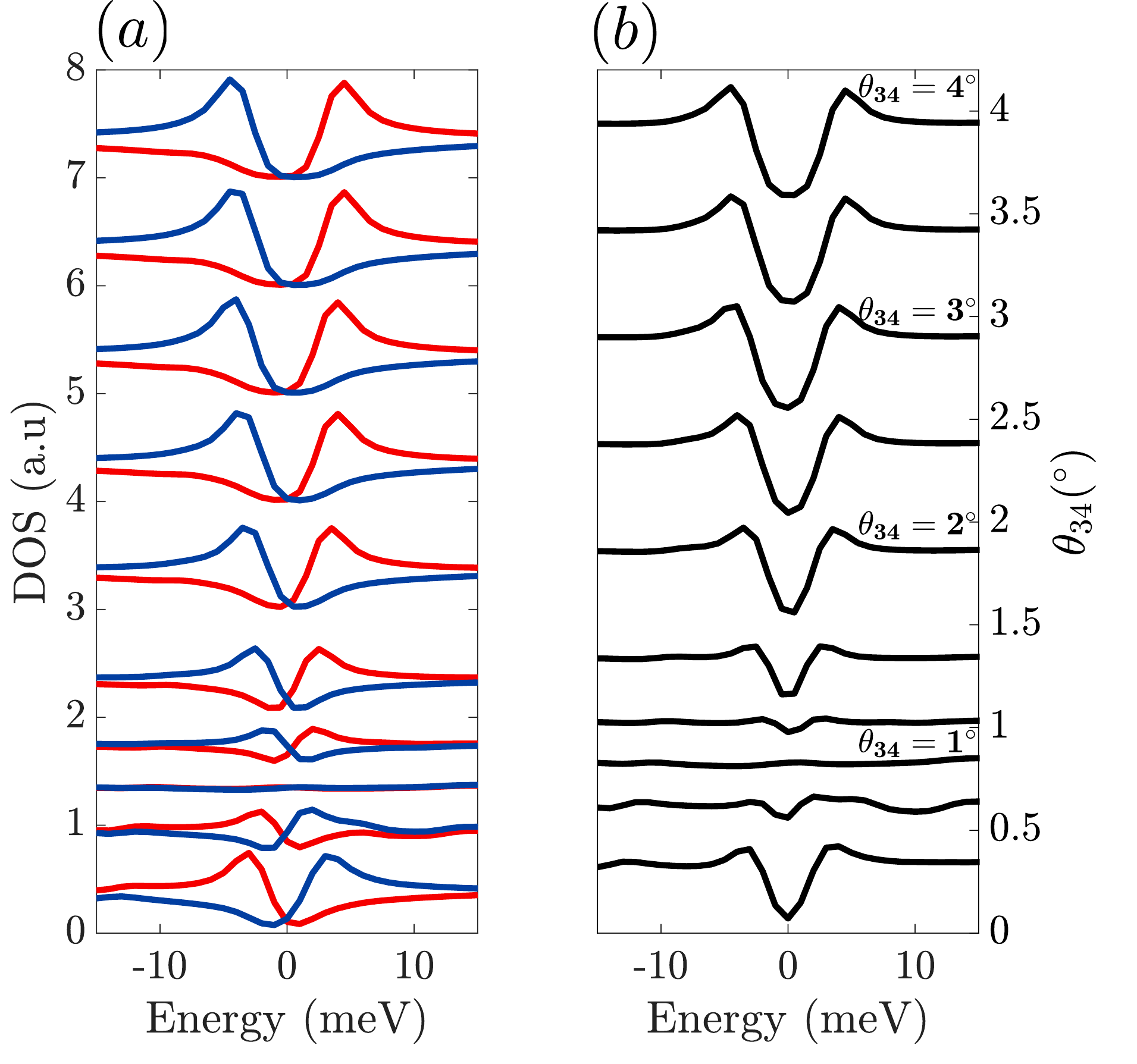}
    \caption{(a) Projected DOS of L1+L2 (red) and L3/L4 (blue) and (b) total DOS near the charge neutrality point for the second twist angle $\theta_{34}$ ranging from $0.2^\circ$(bottom) to $4^\circ$ (top) while the first twist angle $\theta_{12}$ is fixed at $1^\circ$. }
    \label{fig:dos}
\end{figure}

\subsection{Low-energy continuum model}

In previous work we have derived a low-energy momentum space continuum model for the electronic structure of twisted trilayer graphene with two independent twist angles~\cite{zhu2020ttlg}. 
We extend the model to study the quadrilayer heterostructures. 
We note that we do not consider the effect of in-plane relaxation in this work, which forms domain walls in the small-angle limit~\cite{carr2018relaxation,andelkovic2019double,zhu2020relaxation}.
Domain wall formation may affect the polarization of the system, especially the spatial distribution of the ferroelectric domains, and warrants future studies. 
The Hamiltonian can be formally written as the following $4\times 4$ matrix:
\begin{eqnarray}
\mathcal{H} (\vec{q}) = \begin{pmatrix}
H_\mathrm{hBN} & H_{12} & 0 & 0 \\ 
H_{12}^\dagger & H_\mathrm{Gr} & H_{23} & 0 \\
0 & H_{23}^\dagger & H_\mathrm{Gr} & H_{34}  \\ 
0 & 0 & H_{34}^\dagger & H_\mathrm{hBN}
\end{pmatrix},\label{eqn:hamiltonian}
\end{eqnarray}
where $\vec{q}$ is a low-energy momentum around which the Hamiltonian is centered (hereafter referred to as the center site), the diagonal blocks are intralayer terms of the monolayers, and the off-diagonal blocks, $H_{ij}$ are the terms that describe the interlayer interactions. 
Note that the Hamiltonian describes a single valley because the inter-valley separation is much higher in energy than the relevant low-energy degrees of freedom. 
When $\theta_{12}$ and $\theta_{34}$ are not equal, the two valleys are nonequivalent because they are not related by a reciprocal lattice vector. However, we checked that the qualitative behaviors, including the layer polarization and band flatness, of the two valleys are the same and thus we focus on the K valley in the rest of the work.

For the intralayer terms, we take the Dirac Hamiltonian as the monolayer graphene Hamiltonian
\begin{equation}
    H_\mathrm{Gr} (\vec{q}) = - v_F \vec{q} \cdot (\sigma_x, \sigma_y),\label{eqn:monoG}
\end{equation}
where $\sigma_x$ and $\sigma_y$ are Pauli matrices, and $v_F = 0.8 \times 10^6\,\mathrm{cm/s}$ is the Fermi velocity obtained from DFT~\cite{fang2016electronic}.
We adopt the hBN Hamiltonian from \citet{moon2014electronic}:
\begin{equation}
    H_\mathrm{hBN} \approx \begin{pmatrix}
    V_\mathrm{N} & 0 \\ 
    0 & V_\mathrm{B}
    \end{pmatrix},
\end{equation}
where the onsite energies $V_\mathrm{B} = 3.34\, \mathrm{eV}$ and $V_\mathrm{N} = -1.40\, \mathrm{eV}$ capture the sublattice asymmetry of hBN, with the assumption that all the graphene onsite energies are set to $V_\mathrm{C} = 0\,\mathrm{eV}.$ This approximation is justified when the twist angle is small because the hBN potential is far from the low-energy features of interest near the graphene Dirac point. 

The interlayer terms between the graphene -- hBN interface, $H_{12}$ and $H_{34}$, can be derived in a similar way as twisted trilayer graphene by performing a Fourier transform of the real space tight-binding Hamiltonian and taking a low-energy limit~\cite{zhu2020ttlg}. They are given by
\begin{equation}
    H_{\ell,\ell+1} = T^{\ell,\ell+1} (\vec{q}^{(\ell)}, \vec{q}^{(\ell+1)}) = \sum_{n=1}^3 T_{n,\alpha\beta} \delta_{\vec{q}^{(\ell)}-\vec{q}^{(\ell+1)}, -\vec{q}_n^{\ell,\ell+1} },\label{eqn:selection}
\end{equation}
where $\vec{q}^{(\ell)}$ is the low-energy momentum space degree of freedom in layer $\ell$, expanded around $K$ for graphene and $Y$ for hBN monolayers, and the symbols $\{\alpha, \beta\}$ label the sublattice degrees of freedom. We set the first scattering direction as $\vec{q}^{12}_1 = Y_{L1}-K_{L2}$ and $\vec{q}^{34}_1 = K_{L3}-Y_{L4}$, where $K_{L\ell}$ ($Y_{L\ell}$) is the Brillouin zone corner of L$\ell$ for graphene (hBN). Explicitly, $K_\mathrm{L\ell} = \frac{1}{3}(2G_1^{(\ell)} + G_2^{(\ell)})$ for $\ell = 2,3$ and $Y_\mathrm{L\ell} = \frac{1}{3}(2G_1^{(\ell)} + G_2^{(\ell)})$ for $\ell = 1,4$. The other two scattering directions are then generated through rotations $\vec{q}^{\ell,\ell+1}_2 = \mathcal{R}^{-1}(2\pi/3)\vec{q}_1^{\ell,\ell+1}$, $\vec{q}^{\ell,\ell+1}_3 = \mathcal{R}(2\pi/3)\vec{q}_1^{\ell,\ell+1}$, where the rotation $\mathcal{R} (\theta)$ matrix is defined in Eq.~\eqref{eqn:rotation}. The tunneling terms $T_n$ that corresponds to each $\vec{q}^{\ell,\ell+1}_n$ are given as follows, 
\begin{eqnarray}
  T_1 =  u_0 \begin{pmatrix}1 & 1 \\ 
  1 & 1
  \end{pmatrix}, 
  T_2 = u_0 \begin{pmatrix} 1 & \bar{\phi} \\ 
  \phi & 1
  \end{pmatrix}, 
  T_3 =  \bar{T}_2, \label{eqn:tmat}
\end{eqnarray}
where $\bar{z}$ is the complex conjugate of $\bar{z}$, $\phi = \exp (2\pi i/3), \bar{\phi} = \exp (-2\pi i/3),$ and $u_0 = 0.152\,\mathrm{eV}$~\cite{moon2014electronic}.  
Equation~\eqref{eqn:selection} dictates how momentum degrees of freedom between layers $\ell$ and $\ell+1$ are coupled, that is, it prescribes the non-zero matrix elements of $H_{\ell,\ell+1}$. Unlike in bilayer graphene, these degrees of freedom no longer form a simple momentum space lattice due to the interference between the two bilayer moir\'e patterns. Namely, these two moir\'e patterns arise from the lattice-mismatched (and possibly twisted) interfaces between the top hBN and AB bilayer graphene, and between the bottom hBN and AB bilayer graphene. Unless both interfacial twist angles are identical, these two moir\'e lengths are different, and the two moir\'e cells are rotated from each other. As a result, the interference pattern can be dominated by higher-order harmonics and is generally incommensurate even in the continuum limit~\cite{zhu2020relaxation,zhu2020ttlg,doublemoire2020}. 

The off-diagonal term $H_{23}$ represents the interlayer coupling between AB-stacked bilayer graphene, which is approximated to be parabolic:
\begin{eqnarray}
  H_{23} (\vec{q}) = v_F \begin{pmatrix} 0 & \gamma_1 \\ 
  -v_3 (-\nu q_x + i q_y) & 0
  \end{pmatrix},
\end{eqnarray}
where $\nu=\pm 1$ denotes the valley degree of freedom, the parameter $\gamma_1$ represents the band splitting and $v_3$ describes trigonal warping~\cite{mccann2006landau}. We take $\gamma_1 = 0.34$\, eV and $v_3 = 0.051\times 10^6 \,\textrm{m/s}$~\cite{moon2014electronic}.

When a vertical displacement field is applied, each layer has a different external potential energy, which modifies the intralayer terms. With a displacement field $D$, the total potential difference across the 4 layers is $V = 3Dd$, where $d = 3.35$ \AA\ is the interlayer spacing. The potential energy of each layer is then $\Phi_1 = -eV/2, \Phi_2=-eV/6, \Phi_3=eV/6, \Phi_4=eV/2$, and the intralayer term for each layer has an additional term that is $\Phi_\ell \mathbb{1}$ where $\mathbb{1}$ is the $2\times2$ identity matrix. In this way, the positive electric field direction is upward-pointing from L4 to L1. 

\section{Graphene/hBN heterostructures} \label{sec:results}

\subsection{Electronic structures at zero electric  field}\label{sec:no_field}
In the absence of an external electric field, graphene/hBN heterostructures exhibit an intrinsic layer polarization that is dependent on the twist angles. 
Figure~\ref{fig:bands} shows the band structure and DOS of the graphene/hBN heterostructures for three different sets of twist angles. 
The DOS is projected onto the center site $\vec{q}$ along the high symmetry line. Without a twist angle, Fig.~\ref{fig:bands}(a)-(b) shows that neither the valence band nor the conduction band exhibits layer polarization. The low-energy bands are parabolic and the system is gapless, similar to those of AB bilayer graphene.
We compare Fig.~\ref{fig:bands}(a) with the band structure of AB bilayer graphene with aligned hBN on one side in Appendix~\ref{sec:mono_hbn}, in which case the low-energy bands near the charge neutrality point (CNP) exhibit some layer polarization from the hBN layer. 
In Fig.~\ref{fig:bands}(a), note that the low-energy valence and conduction bands cross near the $K$ point. This is analogous to trigonal warping in AB bilayer graphene: when the tunneling between non-dimer sites is included, the parabolic bands are split into four Dirac cones with linear dispersion~\cite{mccann2006landau,rozhkov2016electronic}. Note that in AB bilayer graphene, trigonal warping is very close to the CNP and the saddle point is estimated to be on the order of 1 meV~\cite{mccann2013electronic}.
Here, while we do not include the non-dimer tunneling in the AB bilayer graphene Hamiltonian, the coupling between graphene and hBN serves the same role as remote hoppings and causes trigonal warping. 
When L4 is twisted while L1 is untwisted (Fig.~\ref{fig:bands}(c)-(f)), bands near the CNP become clearly layer polarized: the valence band is L3/L4 polarized whereas the conduction band is L1/L2 polarized. When the two twist angles are equal in magnitude (Fig.~\ref{fig:bands}(g)-(h)), the layer polarization again disappears and the gap closes because the existence of a gap and layer polarization requires the broken inversion symmetry between the top and bottom interfaces. 

Figure~\ref{fig:dos} shows the DOS near the CNP as a function of $\theta_{34}$ while $\theta_{12}$ is kept fixed at $1^\circ$ (note that here L1 and L4 are twisted in the opposite directions). Except for the $\theta_{12} = \theta_{34} = 1^\circ$ case, Fig.~\ref{fig:dos}(a) shows that the DOS is layer polarized. When the inversion symmetry is broken, the system is gapped, and the gap size increases away from $\theta_{12} = \theta_{34}$ and reaches an asymptotic value of $\sim 16\,\mathrm{meV}$ when $\theta_{34} \gtrsim 3^\circ$. Moreover, the layer polarization switches as $\theta_{34}$ crosses $1^\circ$: the valence band is L1/L2 polarized when $\theta_{34} < 1^\circ$ and L3/L4 polarized when $\theta_{34} > 1^\circ$, suggesting that the valence band is polarized by the interface with a larger twist angle.

\begin{figure}[ht!]
    \centering 
    \includegraphics[width=\linewidth]{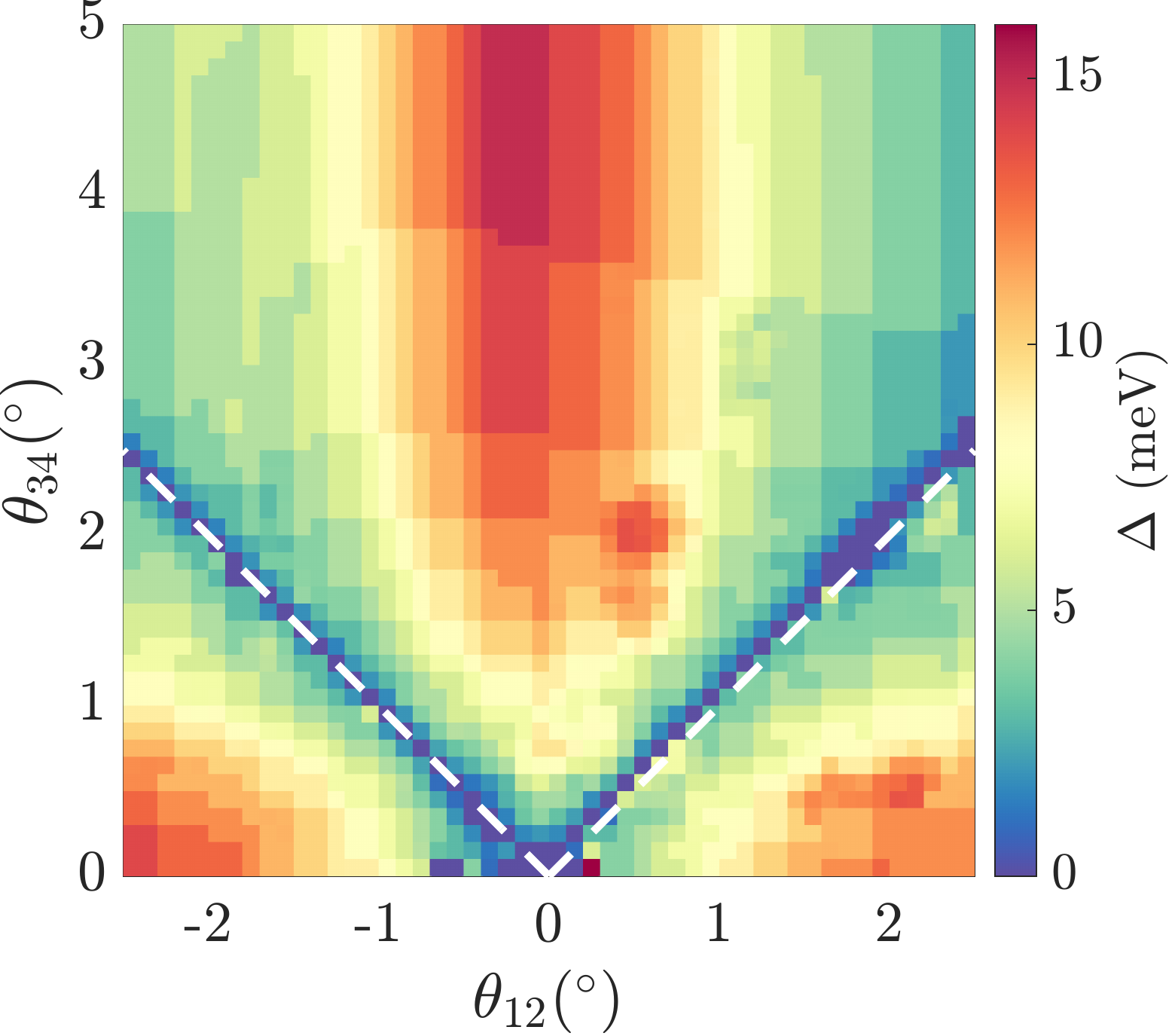}
    \caption{Gap size $\Delta$ between the valence and conduction bands as a function of $\theta_{12}$ and $\theta_{34}$. White dashed lines correspond to $\theta_{34} = \pm \theta_{12}$.}
    \label{fig:gap}
\end{figure}

Figure~\ref{fig:gap} shows the gap size between the valence and conduction bands, $\Delta$, at the CNP in the $(\theta_{12}, \theta_{34})$ phase space. When $\theta_{12} < 0^\circ$, the top and bottom layers are twisted in the same direction. Note that $\Delta$ is symmetric around $\theta_{12} = \pm \theta_{34}$, $\theta_{12} = 0$, and $\theta_{34} = 0$ (the slight asymmetry is due to numerical noise).  When $\theta_{12}= \pm \theta_{34}$, the gap disappears ($\Delta = 0$) because of inversion symmetry. The system is gapped when this inversion symmetry is broken and when at least one of the twist angles is small (|$\theta_{12}| \lesssim 1^\circ$ or |$\theta_{12}| \lesssim 1^\circ$). When both twist angles are greater than $1^\circ$, the gap size $\Delta$ is drastically reduced. This is because when both twist angles are large, the top and bottom hBN layers are essentially decoupled from the AB bilayer graphene layer, and AB bilayer graphene itself has parabolic bands and is gapless. When one of the twist angles is small and the two twist angles are not equal, as the other twist angle increases, the gap size saturates to $\sim 16\,\mathrm{meV}$ as shown in Fig.~\ref{fig:dos}.

\subsection{Wavefunction localization} 

\begin{figure}[ht!]
    \centering
    \includegraphics[width=\linewidth]{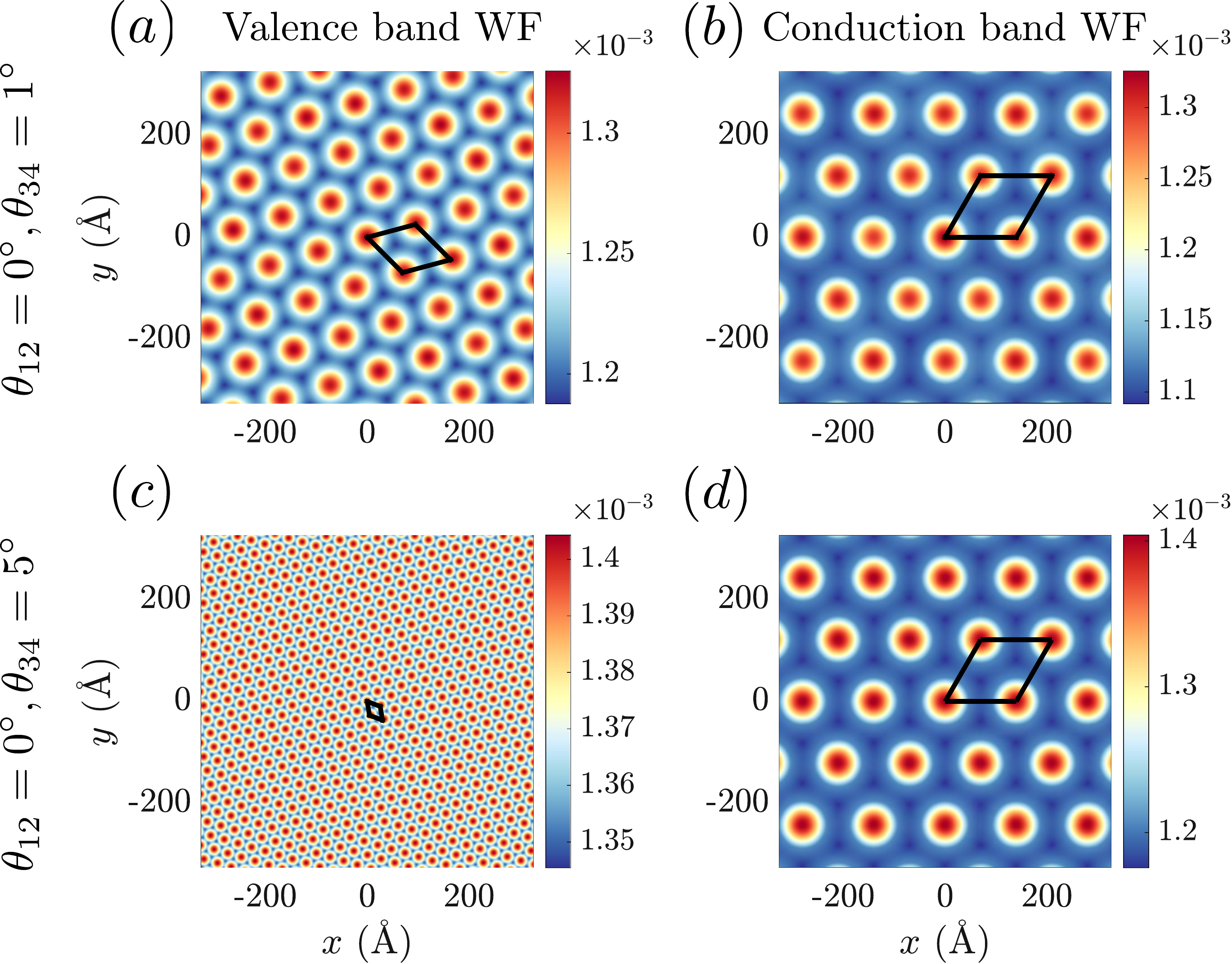}
    \caption{Comparison of the $K$ point wavefunctions $|\Psi_K (\vec{r})|^2$ for (a) valence band and (b) conduction band of $(\theta_{12},\theta_{34})=(0^\circ,1^\circ)$ and (c) valence band and (d) conduction of $(\theta_{12}, \theta_{34})=(0^\circ,5^\circ)$. The black parallelograms in (a) and (c) are the moir\'e supercells that correspond to the $\theta_{34}$ interface, and in (b) and (d) are the moir\'e supercells that correspond to the $\theta_{12}$ interface (in both cases $\theta_{12} = 0^\circ$). }
    \label{fig:wf}
\end{figure}

To obtain the layer-projected real space wavefunction distribution at a position $\vec{r}$, we perform an inverse Fourier transform by summing over the wavefunction weights that correspond to each momentum degree of freedom $\vec{q}^{(\ell)}$ on layer $l$:
\begin{eqnarray}
  \Psi^\ell_{n, \vec{q}} (\vec{r}) = \sum_{\alpha = A,B} \sum_{\vec{q}^{(\ell)}} \psi_{n, \vec{q}, \alpha} (\vec{q}^{(\ell)}) e^{- i \vec{q}^{(\ell)}\cdot \vec{r} }, 
\end{eqnarray}
where $n$ is the band index, $\vec{q}$ is the center site of the Hamiltonian, and $\psi_{n,\vec{q},\alpha} (\vec{q}^{(\ell)}) $ is the wavefunction weight that corresponds to momentum $\vec{q}^{(\ell)}$ and sublattice $\alpha$. We sum over a total number of 5840 momentum degrees of freedom in the basis (values of $\vec{q}^{(\ell)}$).
Figure~\ref{fig:wf} shows the $K$-point ($\vec{q} =$ $K$) valence and conduction band wavefunction distribution in real space for $(\theta_{12},\theta_{34}) =(0^\circ, 1^\circ)$ and $(\theta_{12},\theta_{34}) = (0^\circ, 5^\circ)$. In both cases, valence band wavefunctions localize at the moir\'e supercell that corresponds to $\theta_{34}$ whereas the conduction band wavefunctions localize at the $0^\circ$ supercell from the lattice mismatch. This is consistent with the band layer polarization in Fig.~\ref{fig:bands} (c) and (e) -- since the valence bands are L3/L4-polarized, the wavefunctions are also expected to have the length scale of the L3/L4 supercell, and vice versa for the conduction bands. We could consider the wavefunction distribution as the summation of localized electrons at the moir\'e scale and a uniform background. The scale of the localization depends on the interface while the uniform background can be attributed to band dispersion. 

\begin{figure*}[ht!]
    \centering
    \includegraphics[width=\linewidth]{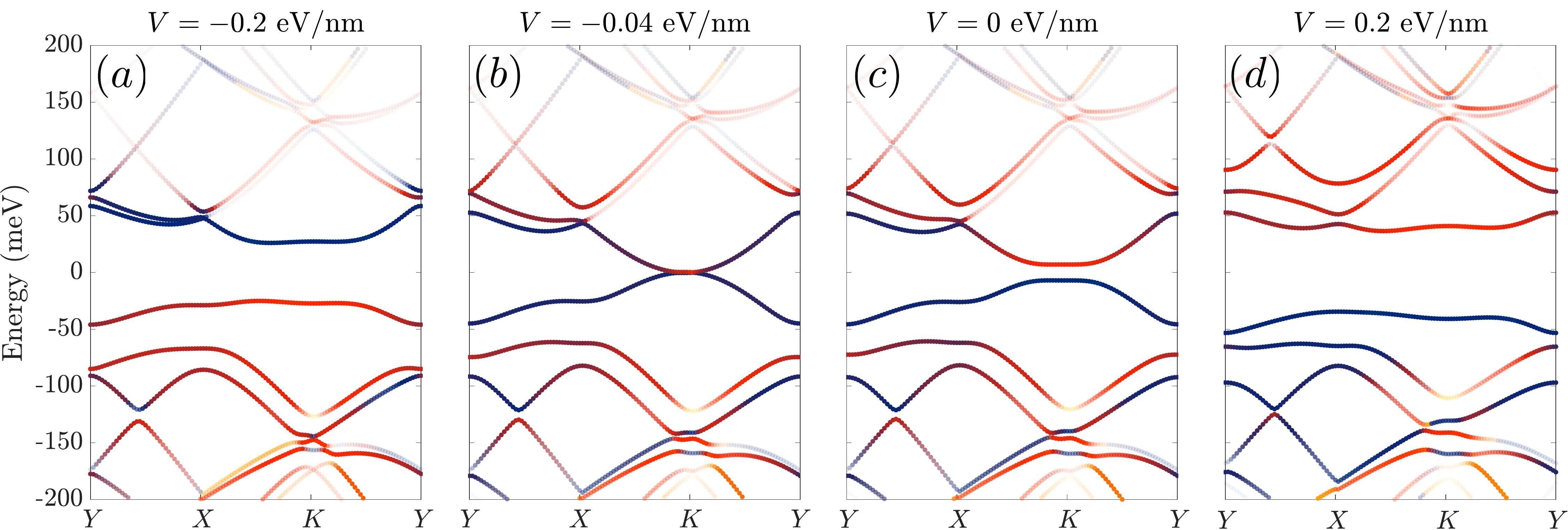}
    \caption{Band structures for graphene/hBN heterostructures with $(\theta_{12},\theta_{34})=(-0.1^\circ, 15^\circ)$ with (a) $V=-0.2\,\mathrm{eV/nm}$, (b) $V=-0.04\,\mathrm{eV/nm}$, (c) $V=0\,\mathrm{eV/nm}$, and (d) $V = 0.2 \, \mathrm{eV/nm}$. The same color scheme as in Fig.~\ref{fig:bands} is used; the high symmetry line corresponds to the red dashed lines in Fig.~\ref{fig:reciprocal}. }
    \label{fig:bands_efield}
\end{figure*}

\begin{figure}[ht!]
 \centering
    \includegraphics[width=\linewidth]{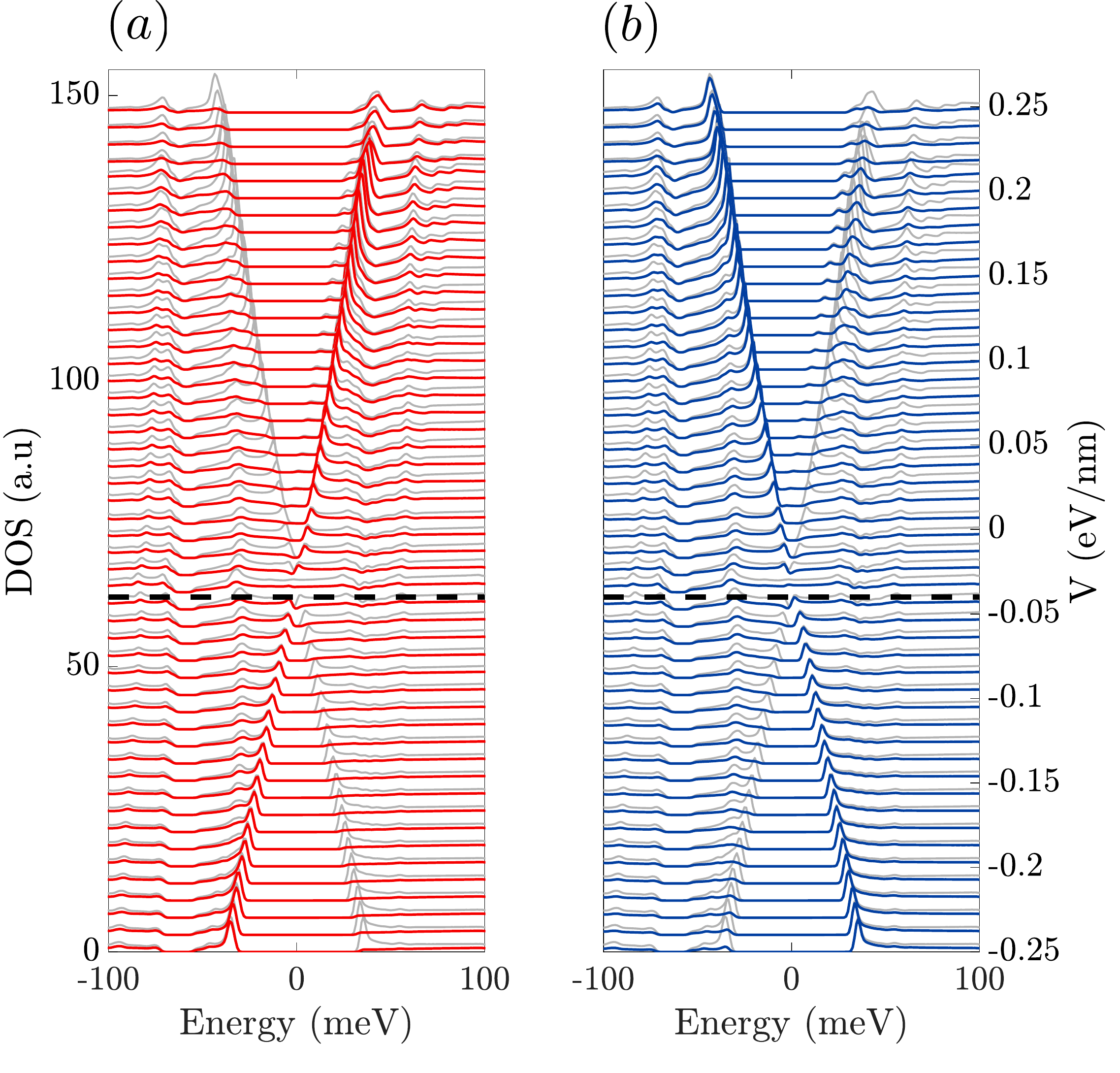}
    \caption{Projected DOS onto (a) L1 + L2 and (b) L3 + L4 as a function of external electric fields for $(\theta_{12}, \theta_{34}) = (-0.1^\circ, 15^\circ)$. In both subplots, the gray lines are the corresponding total density of states. The y-axis on the left labels the magnitude and direction of the applied electric field, from -0.3 eV/nm to 0.25 eV/nm with a 0.01 eV/nm increment. The black dashed line is $V = -0.04\,\mathrm{eV/nm}$, which is the critical electric field where the gap closes and polarization switches. Note that the same normalization constant is used for panels (a) and (b).  }
    \label{fig:dos_efields}
\end{figure}

\begin{figure}[ht!]
    \centering
    \includegraphics[width=\linewidth]{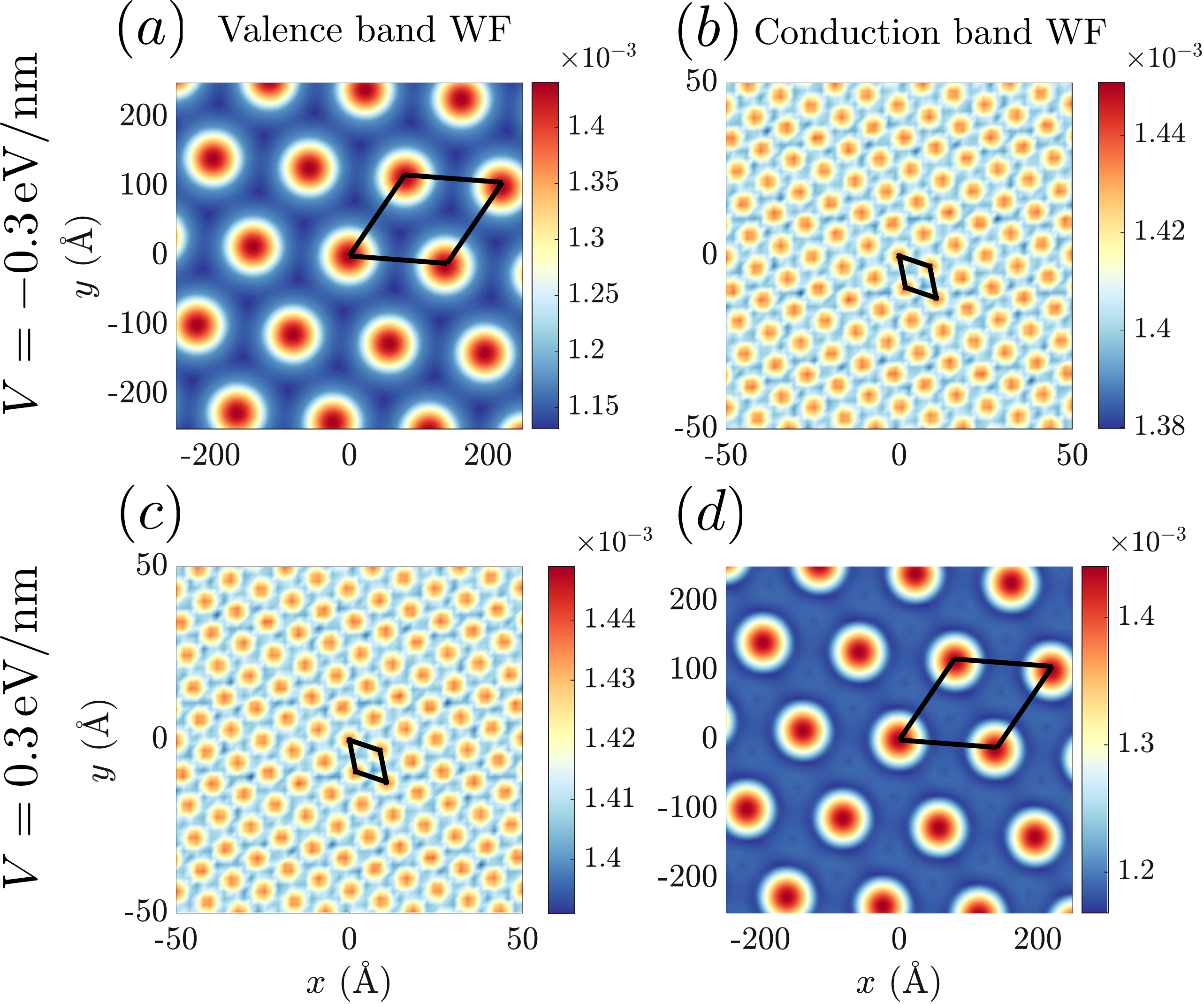}
    \caption{Comparison between the $K$-point wavefunctions $|\Psi_K (\vec{r})|^2$ for $(\theta_{12}, \theta_{34}) = (-0.1^\circ, 15^\circ)$. (a) Valence and (b) conduction band wavefunctions with $V = -0.3 \, \mathrm{eV/nm}$; (c) Valence and (d) conduction band wavefunctions with $V = 0.3 \, \mathrm{eV/nm}$. The black parallelograms in (a) and (d) are the moir\'e supercell that corresponds to $|\theta_{12}| = 0.1^\circ$, and in (b) and (d) are the moir\'e that corresponds to $\theta_{34}=15^\circ.$}
    \label{fig:wf_efield}
\end{figure}

\subsection{Electric field dependence} 
The layer polarization of the conduction and valence bands can be tuned by an out-of-plane displacement field. Figure~\ref{fig:bands_efield} shows the band structure for ($\theta_{12},\theta_{23})=(-0.1^\circ, 15^\circ)$ for four different values of the applied electric field. 
In absence of an external electric field (Fig.~\ref{fig:bands_efield}(c)), as shown in Section~\ref{sec:no_field}, the valence band is polarized to the interface with the larger twist angle, which is L3/L4. The polarization gives rise to a positive or upward-pointing intrinsic electric field, from the bottom L4 to the top L1. When a negative electric field is applied, there is a critical value of which the applied electric field cancels out the intrinsic polarization. In this case, the critical electric field is $V_\mathrm{c} = -0.04\,\mathrm{eV/nm}$ (Fig.~\ref{fig:bands_efield} (b)). In general, the value of the critical electric field depends on the single-particle gap size $\Delta$ (Fig.~\ref{fig:gap}), and the larger $\Delta$ is, the larger the $V_c$ is. As the negative electric field increases, the polarization reverses (Fig.~\ref{fig:bands_efield}(a)) and the valence band becomes L1/L2 polarized while the conduction band is L3/L4 polarized, dominated by the direction of the external electric field. As the magnitude of the electric field becomes large, for both positive and negative electric fields, the valence and conduction bands flatten significantly and the gap increases
(Fig.~\ref{fig:bands_efield}(a)(d)). 
Figure~\ref{fig:dos_efields} shows the DOS of $(\theta_{12}, \theta_{34}) = (-0.1^\circ, 15^\circ)$ as a function of displacement fields. The projected DOS onto the L1/L2 and L3/L4 interfaces in Fig.~\ref{fig:dos_efields} clearly shows the polarization reversal at the critical electric field $-0.04\,\mathrm{eV/nm}$. Both the gap size and the DOS maximum smoothly increase as the applied electric field increases. Moreover, we note that the DOS has a particle-hole asymmetry at large electric fields. For example, at a large positive electric field (say $0.25\,\mathrm{eV/nm}$), the valence band peak clearly has a larger magnitude than the conduction band.

In addition to the band structures and DOS, the external electric field also changes the wavefunction localization. Figure~\ref{fig:wf_efield} compares the valence and conduction band wavefunctions with positive and negative applied electric fields. When $V < V_c$ (Fig.~\ref{fig:wf_efield}(a)-(b)), the valence bands have the length scale of the $|\theta_{12}|=0.1^\circ$ moir\'e supercell, while the conduction band wavefunctions localize at the much smaller $15^\circ$ moir\'e supercell. This agrees with the layer polarization reversal by the negative electric field as shown in Figs.~\ref{fig:bands_efield} and \ref{fig:dos_efields}. 
When $V > V_c$, Fig.~\ref{fig:wf_efield}(c)-(d) shows that the wavefunctions localize at the opposite length scales as in (b)-(d), which is also the same scales as the zero-electric field case because the applied electric field has the same direction as the intrinsic electric field, and thus there is no polarization reversal. 
Note that we do not include the dielectric permittivity of the hBN and graphene when modeling the effect of the electric field. Including the dielectric effect changes the critical electric field where the gap closes but is not expected to affect the qualitative behaviors discussed in this section including the gap closing, polarization reversal, and band flattening.

\section{Quadrilayer graphene heterostructures}~\label{sec:graphene}
We could employ the low-energy continuum model to study quadrilayer graphene heterostructures, with AB bilayer graphene encapsulated by top and bottom graphene layers. The intralayer terms for monolayer hBN, $H_\mathrm{hBN}$ in Eq.~\eqref{eqn:hamiltonian} are replaced by rotated Dirac Hamiltonian, $H_\mathrm{Gr}$, which represents a slight modification to Eq.~\eqref{eqn:monoG} and is defined as follows
\begin{equation}
    H_\mathrm{Gr} (\vec{q}) = - v_F \vec{q} \cdot (\sigma_x^{\theta_\ell}, \sigma_y^{\theta_\ell}),\label{eqn:monoG_rotated}
\end{equation}
where $\sigma_x^{\theta_\ell} = \sigma_x\cos \theta_\ell- \sigma_y \sin \theta_\ell, \sigma_y^{\theta_\ell} = \sigma_x \sin \theta_\ell + \sigma_y \cos \theta_\ell$ are rotated Pauli matrices with $\theta_1 = \theta_{12}$, $\theta_{2} = 0$, $\theta_3 = -\theta_{34}$.
The interlayer interactions between the top and middle layers, $H_{12}$ and $H_{34}$, have the same form as Eq.~\eqref{eqn:tmat}, but the three scattering vectors $\vec{q}^{\ell,\ell+1}_j$ for $j=1,2,3, \ell = 1,3$ are given by $\vec{q}^{\ell,\ell+1}_1 = K_{L\ell}-K_{L\ell+1}$, $\vec{q}^{\ell,\ell+1}_2 = \mathcal{R}^{-1}(2\pi/3)\vec{q}_1^{\ell,\ell+1}$, and $\vec{q}^{\ell,\ell+1}_3 = \mathcal{R}(2\pi/3)\vec{q}_1^{\ell,\ell+1}$. In addition, the scattering matrices $T_j$ that correspond to each $\vec{q}^{\ell,\ell+1}_j$ are defined in the same way, 
\begin{equation}
    T_1 = \begin{pmatrix} \omega_0 & \omega_1 \\ 
    \omega_1 & \omega_0 
    \end{pmatrix}, \quad
    T_2 = \begin{pmatrix} \omega_0 & \omega_1 \bar{\phi} \\ 
    \omega_1 \phi & \omega_0
    \end{pmatrix}, \quad
    T_3 = \bar{T}_2,
\end{equation}
where $\omega_0 = 0.07\,\mathrm{eV}$ and $\omega_1 = 0.11\,\mathrm{eV/nm}$ are the interlayer hopping parameters between AA and AB stackings respectively and the difference between $\omega_0$ and $\omega_1$ accounts for the out-of-plane relaxation~\cite{nam2017lattice,carr2019exact,fang2019angle}.

\begin{figure}[ht!]
    \centering
    \includegraphics[width=\linewidth]{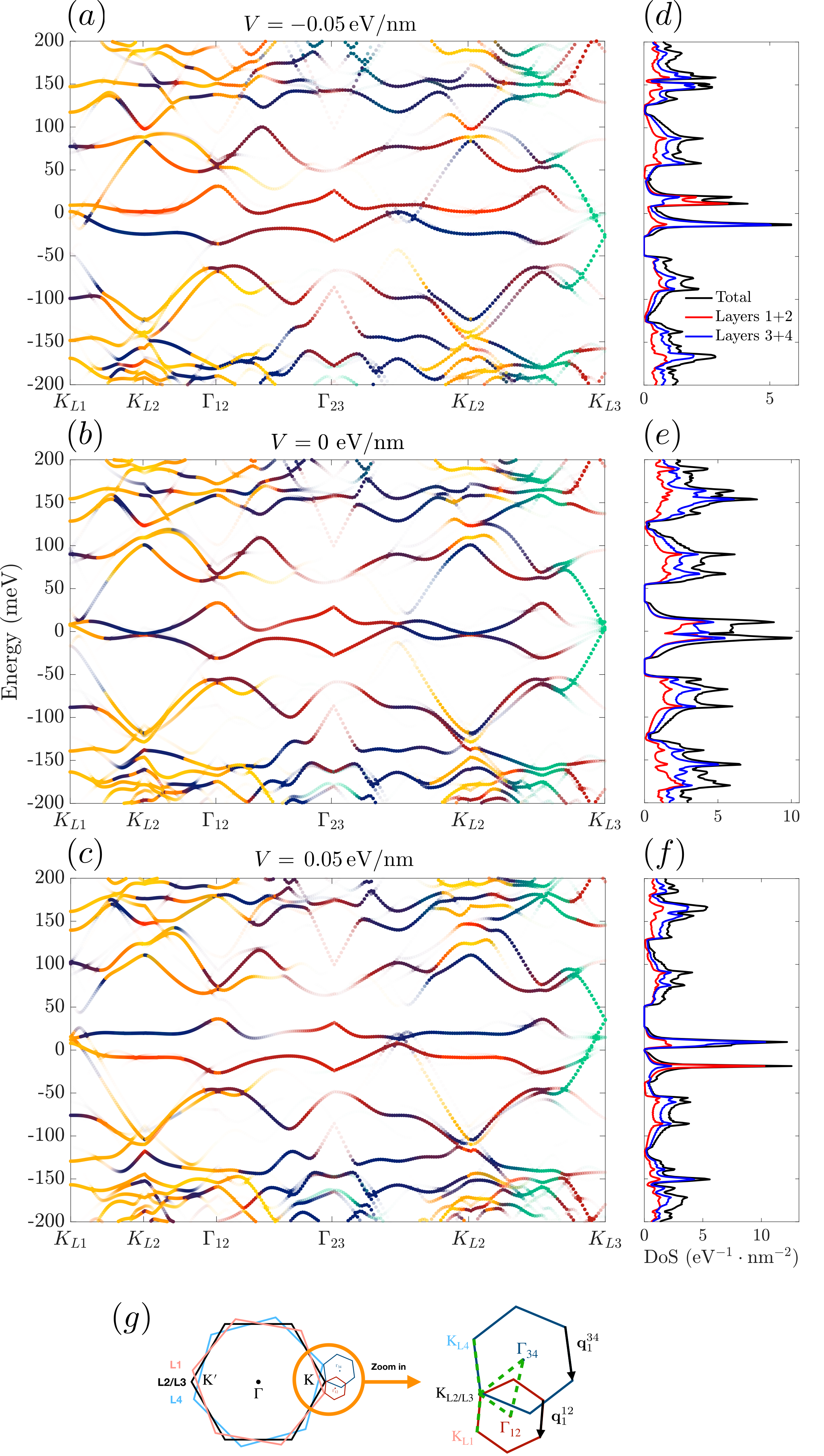}
    \caption{Electronic structures for AB bilayer graphene encapsulated by top and bottom graphene layers with $\theta_{12}, \theta_{34}=(1.6^\circ, 3.0^\circ)$. The left panels shows the band structure along the high symmetry line in (g), with (a) $V = -0.05 \,\mathrm{eV/nm}$, (b) $V = 0 \,\mathrm{eV/nm}$, and (c) $V = 0.05 \,\mathrm{eV/nm}$ respectively. Colors represent the projected weights along the high-symmetry line cut with the same color scheme as shown in Fig.~\ref{fig:bands}(a). (d), (e), (f) The corresponding DOS of (a), (b), and (c). (g) Reciprocal space of the quadrilayer graphene heterostructures, where the light red, black, and, light blue hexagons are monolayer graphene Brillouin zones. On the left, the dark red and blue hexagons are the moir\'e Brillouin zones of L1/L2 and L3/L4 respectively. }
    \label{fig:graphene bands}
\end{figure}

Figure~\ref{fig:graphene bands} shows the electronic structure of the quadrilayer graphene heterostructures with $(\theta_{12}, \theta_{34}) = (1.6^\circ, 3.0^\circ)$. Unlike the hBN/graphene heterostructures, the quadrilayer graphene system exhibits no significant layer polarization or band gap in absence of an external electric field.
The application of the electric field introduces layer polarization to the bands near the CNP, and the induced layer polarization agrees with the applied electric field direction (Fig.~\ref{fig:graphene bands}(a)-(f)). In addition, for certain values of applied electric fields, (for this twist angle combination when $V = 0.05 \,\mathrm{eV/nm}$), the low-energy bands near the CNP become significantly flatter and the symmetry between the valence and conduction is also broken. However, the system remains gapless under the applied electric field.

\section{Discussion and Conclusion} ~\label{sec:conclusion} 
In the experiment by ~\citet{zheng2020}, the significant feature associated with the ferroelectric switching is an LSAS, namely, under certain conditions, the top (bottom) layer can completely screen the electric field from the top (bottom) while the electric field can still penetrate the bottom (top) layer. Such an asymmetric screening scenario points to vastly different electronic structures associated with different layers. Our calculations show that the twist angle difference between the top BN/graphene interface and the bottom BN/graphene interface could lead to layer-asymmetric electronic structures, which provides the starting point to understanding such unusual LSAS behaviors. In particular, if one interface (top) has a longer moir\'e wavelength and the other interface (bottom) has a shorter moir\'e interface (Figs.~\ref{fig:dos_efields} and \ref{fig:wf_efield}), under certain conditions the top layer will feature electronic structures with a large DOS and localized wavefunctions but not the bottom layer. This can support a scenario where the top layer may feature a stronger correlation effect and unique screening properties.

Based on the information from the band structure, we could provide a rough estimate of the intrinsic polarization as follows. 
If we fill the conduction band with one electron per the moir\'e cell that corresponds to the 0$^\circ$ interface, according to, for instance, Fig.~\ref{fig:bands}(e) and the weight distribution on different layers, effectively all the weights are concentrated at the L1/L2 interface (99.7\%) and we can treat the electron to be purely localized at the L1/L2 interface. 
As a result, the dipole moment can be calculated as $p = 3ed$, where $e$ is the electron charge and $d=3.35$ A is the interlayer distance. The electric polarization is then $P=p/(3A_\mathrm{m}d)$, where $A_\mathrm{m}$ is the moir\'e superlattice area and $A_\mathrm{m}=1.69\times 10^{-12}\, \mathrm{cm}^{2}$ for the $0^\circ$ interface. Using these, we estimate $P = 0.095$ C/cm$^{-2}$. Despite being a crude estimate that ignores the dielectric environment, this is on the same order of magnitude with the experimentally measured polarization by \citet{zheng2020}, which is 0.12 C/cm$^{-2}$. 

In addition to the hBN/graphene heterostructure, we showed that the low-energy band structure of the quadrilayer graphene heterostructures can be tuned by external electric fields in a similar way. While at the zero electric field the low-energy moir\'e bands do not exhibit moir\'e induced intrinsic polarization, an external electric field not only alters the band flatness but also breaks the symmetry between the valence and conduction bands. If the observed ferroelectricity in hBN-sandwiched AB bilayer graphene has an electronic origin, we expect the quadrilayer graphene system to exhibit similar hysteretic behavior in transport measurements.

\acknowledgments{We thank Zhiren Zheng, Suyang Xu, and Xueqiao Wang for helpful discussions. Z.Z. and E.K. acknowledge the STC Center for Integrated Quantum Materials, NSF Grant No. DMR-1922172, ARO MURI Grant No. W911NF-2120147, and NSF DMREF Grant No. 1922165, and Simons Foundation, Award No. 896626.
S.C.
is supported by the National Science Foundation under grant No. OIA-1921199. Q.M. is supported by the Center for the Advancement of Topological Semimetals, an Energy Frontier Research Center funded by the US Department of Energy, Office of Science, through the Ames Laboratory under contract DE-AC02-07CH11358. Q.M. also acknowledges support from the CIFAR Azrieli Global Scholars program.
Calculations were performed on the Cannon cluster supported by the FAS Division of Science, Research Computing Group at Harvard University, and National Energy Research Scientific Computing Center (NERSC), a U.S. Department of Energy Office of Science User Facility located at Lawrence Berkeley National Laboratory, operated under Contract No. DE-AC02-05CH11231 using NERSC award BES-ERCAP0020773.}

\appendix

\section{Computational Details of DOS Calculations} 
To properly normalize the DOS, we first calculate the DOS of the intralayer Hamiltonian only, which reduces to four independent monolayers. Near the charge-neutrality point, the DOS per $\mathrm{eV}$ per $\mathrm{nm}^2$ is given by~\cite{castro_neto2009electronic}
\begin{equation}
    \mathcal{D} (\epsilon) = \frac{2D}{\pi}\frac{|\epsilon|}{v_F^2}, 
    \label{eqn:dos_slope}
\end{equation}
where the prefactor includes a factor 4 from spin and valley degeneracies and $D$ is system-dependent. For hBN/graphene heterostructures, $D = 2$ because only two graphene monolayers contribute to the low-energy DOS and the monolayer hBN is estimated to be a constant potential that is high in energy. For quadrilayer graphene heterostructures, $D=4$ to account for the contribution from all 4 layers. 
We then obtain a normalization constant by fixing the prefactor to the expected slope given in Eq.~(\ref{eqn:dos_slope}) and use the same constant for the DOS of the full Hamiltonian.

The density of states (DOS) is obtained by discretizing the moir\'e Brillouin zone between L1 and L2 by $61 \times 61$. We adapt the Gaussian smearing full-width-half-maximum $\sigma$ based on $\theta_{12}$ because the moir\'e Brillouin zone area being sampled increases as the twist angle increases. For graphene/hBN heterostructures, $\theta_{12} < 0.8^\circ, \sigma = 1.7 \,\mathrm{meV}$, $\theta_{12} < 2^\circ, \sigma = 2.3 \,\mathrm{meV}$, $\theta_{12} \geqslant 2^\circ$, $\sigma = 2.7\,\mathrm{meV}$. For quadrilayer graphene heterostructure with $\theta_{12}=1.6^\circ, \theta_{34} = 3^\circ$, we choose $\sigma = 1.2\,\mathrm{meV}$. Each point $\vec{q}$ on this BZ grid is then expanded into a truncated momentum-space Hamiltonian of dimension $2028 \times 2028$, which is equivalent to truncating the basis to the second shell of the monolayer lattice vectors before taking the low-energy expansion (namely, we constrain the magnitude of $\vec{k}^{(\ell)} = \vec{q}^{(\ell)} + K_{K\ell/Y_\ell}$). 
We confirmed that the features of interest have converged with respect to the momentum-basis cutoff radius. 
% theta<0.8, w_inv = 200 (1.7 meV), theta<2, w_inv = 150 (2.3 meV) , else w_inv = 130 (2.7 meV)

\begin{figure}[ht!]
    \centering
    \includegraphics[width=0.8\linewidth]{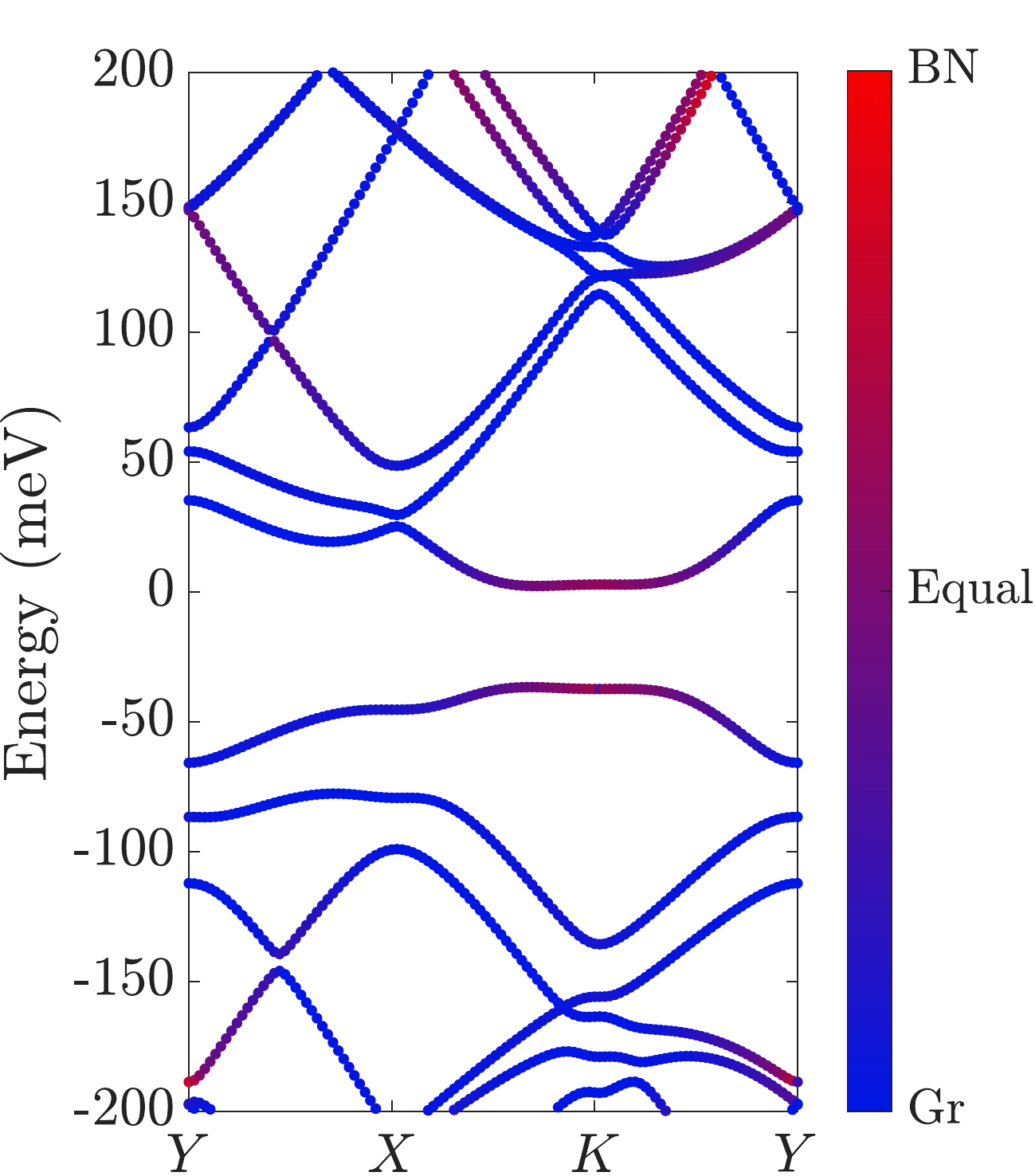}
    \caption{Electronic band structure of AB bilayer graphene on top of monolayer hBN with no twist. Colors are projected onto the center site along the high symmetry line, with red being purely on the hBN layer, blue being purely on the graphene layer}
    \label{fig:monoG_hbn}
\end{figure}

\section{AB bilayer graphene + monolayer hBN electronic structure} ~\label{sec:mono_hbn}
Figure~\ref{fig:monoG_hbn} shows the electronic band structure of AB bilayer graphene on top of a monolayer hBN layer. The band structure agrees with the results in \citet{moon2014electronic}. Unlike in the hBN-sandwiched AB bilayer graphene without a twist (Fig.~\ref{fig:bands}(a)), this trilayer system is gapped because of the broken inversion symmetry due to the hBN monolayer. The low-energy bands have hBN characteristics because of the net potential from the hBN -- we approximate the hBN potential to be constant, $V_B$ and $V_N$, leading to a net constant potential of $\sim 2\, \mathrm{eV}$ and thus polarizes the low-energy bands. In contrast, the net hBN potential is canceled out in the hBN-sandwiched graphene heterostructures. Therefore, in the quadrilayer graphene/hBN heterostructures, in absence of a twist angle, we expect the system to be layer unpolarized, and the twist angle introduces moir\'e-induced layer polarization.

\bibliography{references}
\end{document}